\documentstyle[sprocl]{article}

\input{psfig}
\def\be{\begin{equation}}
\def\ee{\end{equation}}
\def\bea{\begin{eqnarray}}
\def\eea{\end{eqnarray}}

\begin{document}

\rightline{OITS 632}
\rightline{August 1997}

\title{THE PHOTON SPECTRUM IN QUARKONIA DECAYS\footnote{To appear 
in the Proceedings of the Conference PHOTON97, 10-15 May 1997, 
Egmond-aan-Zee, The Netherlands.}}

\author{ F. HAUTMANN }

\address{Institute of Theoretical Science, University of Oregon \\
Eugene, OR 97403 }

\maketitle\abstracts{
I discuss  recent theoretical results  
 for inclusive decays of quarkonia into a photon plus hadrons, and 
summarize the status of perturbative calculations. }

Radiative decays of 
heavy quark bound states 
have been investigated in the framework of  perturbative QCD    
and  
have been used 
to measure   
the QCD coupling at scales of the order of the heavy quark mass 
(for  recent 
studies see Refs.~[1-3],  and   
references therein).  
Away from the boundaries of the phase space, 
the inclusive spectra are 
expected to be 
well described by  
first or second order 
perturbation theory. Near the phase space boundaries, 
effects associated with non-perturbative contributions  
and with high orders in perturbation theory   
are expected to 
become 
 important. 
In this paper I 
focus on two of these effects. 
 First, 
  I describe  the role of 
the fragmentation of gluons and light quarks 
into a photon. 
This 
is 
relevant to the shape of the inclusive photon spectrum 
at small $z$, where $z =  2 \, E_\gamma / M $ is 
the energy fraction carried by the photon, with $M$ being 
the quarkonium mass. 
 Second,  I consider 
  the opposite end of the spectrum, $z \to 1$.   
I discuss the 
resummation of soft gluons 
in this region and its physical  implications.

Consider 
the  decay 
of a $^{3} S_1$ quarkonium state  into a prompt photon plus hadrons: 
\begin{equation}
\label{reaction}
X_{ Q {\bar Q}} \to \gamma + {\mbox {hadrons}} 
\hspace*{ 0.6 cm} . 
\end{equation}
It was pointed out in Ref.~[4] that this reaction    
receives contributions  at leading logarithmic 
order (LO)  
both when the photon is coupled to highly virtual processes 
(``direct term") 
and 
 also when the 
 photon is produced by collinear emission from 
 light quarks  (``fragmentation term"). As a consequence of the 
 factorization theorem for collinear mass singularities, the 
decay width $ d \Gamma / d z$ has the structure  
\begin{equation}
\label{factzation}
{{ d \Gamma} \over {  d z }} (z, M) = 
C_\gamma \left( z , \alpha_s(M) \right) + 
\sum_{a = q , {\bar q} , g} 
\, \int_z^1 \, {{ d x } \over x} \, 
 C_a \left( x , \alpha_s (M) \right) \, 
 D_{ a  \gamma} 
\left( {z \over x} , M \right)  
\hspace*{ 0.3 cm} . 
\end{equation} 
  The first term in the right hand side 
 denotes  the direct contribution, while 
 the second term denotes 
the fragmentation contribution. 
The functions  
$C_A$, with 
$A = \gamma, q, {\bar q}, g$,  are 
short-distance coefficient functions, calculable as 
power series expansions in $ \alpha_s $:  
\begin{equation}
\label{hardcoe}
C_A = C_A^{(0)} + 
\alpha_s  \,    
C_A^{(1)}
+ \cdots 
\;\;.  
\end{equation}
The functions 
$ D_{ a  \gamma}$ are the fragmentation functions 
 for $ a \to \gamma + X$, 
satisfying evolution equations of Altarelli-Parisi type. 
Through these functions, the process 
 is 
sensitive to long-distance physics. 
In general, both the $C$'s and $D$'s depend on a factorization scale 
$\mu$.    
 The separation between direct and fragmentation components 
depends on the choice of this scale.  
For simplicity, 
in Eq.~(\ref{factzation}) 
 the factorization scale $\mu$ 
 and 
 the  
renormalization scale that appears in the running coupling 
have been set equal to $M$.   

\begin{figure}
\psfig{figure=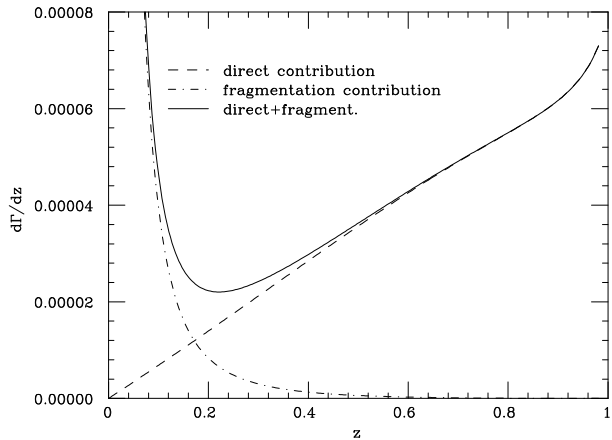,height=2.9in}
\caption{ 
The photon spectrum in   $\Upsilon$  decay   
to leading order (LO). The solid curve is the full LO result. 
The dashed curve is the contribution from the direct term, while 
the dot-dashed curve is the contribution from the 
fragmentation term. 
 We use units $4 \, \pi \, \psi_0^2/M^2=1$, with $\psi_0$ being the 
bound state wave function at the origin, 
 and 
$\alpha_s=0.2$. 
\label{dirfraplots}}
\end{figure}

The direct term to LO, ${C}_{\gamma }^{(0)}$, 
stemming from the decay $X_{ Q {\bar Q}} \to \gamma g g $,   
was calculated 
  in Ref.~[5]. 
It is shown by the dashed curve in Fig.~1 
for the case of $\Upsilon (1 S)$. 
 The fragmentation terms  to LO, 
${C}_{a }^{(0)} \otimes D_{a \gamma}$, 
 were calculated in Ref.~[4].   
To this order, 
 the  fragmentation contribution comes entirely from 
 the gluon channel  
($a = g$), as  
coefficients   in the 
quark channels  
($a = q , {\bar q}$)  
vanish. 
This contribution  
is shown by the 
dot-dashed curve in Fig.~1, where 
we use 
the 
set of 
fragmentation functions 
given in   
Ref.~[6].  
The solid curve in Fig.~1 is given by the sum of the 
direct and fragmentation terms, and provides 
the full LO result.

We see that the fragmentation component is 
suppressed with respect to the direct component 
if $z$ is sufficiently large. As $z$ decreases,  
fragmentation becomes important. In particular,  
to this order 
in perturbation theory one finds the behavior 
in $1 / z$ for $z \to 0$ 
characteristic of the soft 
 bremsstrahlung spectrum.

Fig.~1 suggests that, if one wishes to  
 analyze the inclusive photon data 
in $\Upsilon$ decay 
by using  theoretical formulas 
 based on direct production only, 
one should restrict oneself to 
the region above some minimum value $z_{\rm {min}}$. 
However, 
the question of how big this value should be is also influenced 
by the size of the next-to-leading  terms.  
 I will comment on this 
 below.  
Alternatively, and more interestingly, Fig.~1 suggests that, 
if 
the experimental errors on 
the measurement of the spectrum at relatively low $z$ 
  were  
reduced with respect to their present very large values~\cite{cleo},  
one could use this process to     
   perform a test of the full QCD prediction,  
 and to learn  about  gluon fragmentation  
into 
photons.

The theory  
 has not  been fully developed 
to the next-to-leading order (NLO)  
yet. 
 The  NLO coefficients  
$ {C}_{A}^{(1)}$ are not known at present. Only the result 
 of a calculation for the integral  
$ \int_0^1 dz\,{C}_{\gamma}^{(1)}(z)$ is available~\cite{ML}.   
 Besides, there have been 
studies of 
  $ {C}_{\gamma}(z)$ 
 based on modeling~\cite{Field,Photia}  certain classes of 
higher order corrections. A calculation 
of ${C}_{\gamma}^{(1)}(z)$ is in progress~\cite{krae}. 
The full calculation of the  coefficients $ {C}_{A}^{(1)}$ 
would be  important. It could be combined with the results 
of the next-to-leading 
evolution of fragmentation functions~\cite{nlofra} 
to give a consistent NLO treatment of radiative $\Upsilon$ decays.  
In particular, 
 note that in NLO quark fragmentation starts to 
contribute. Quark fragmentation 
functions are much harder than gluon fragmentation functions. 
As a result,  the value of the 
cut $z_{\rm {min}}$ introduced above 
might be pushed 
significantly to the right 
with respect to the value that can be read 
from Fig.~1.  

The second aspect that I 
 discuss 
in this paper 
concerns  
the region near the endpoint $z = 1$ of the 
spectrum. 
 This region 
 is influenced by 
non-perturbative QCD effects.  
 The shape of the spectrum is expected to become 
very sensitive to hadronization corrections 
 as $z \to 1$. 
A way of dealing  with such effects  was proposed in  the   
Monte Carlo study of 
  Ref.~[8].     
This calculation uses 
a parton-shower description for the gluon radiation 
 in higher orders,  and 
an independent-fragmentation 
model for hadronization. 
A different approach  
may be found  in  
  Refs.~[2,12], based on    parametrizing   
  non-perturbative contributions   
in terms of an effective gluon mass~\cite{pape}. 

Further sensitivity to  nonperturbative parameters 
may 
enter 
through 
 higher  Fock states in the quarkonium.
The role of 
these states has been studied recently~\cite{rothwise}. 
Away from $z =1$, the color-singlet Fock state dominates the 
 decay,  
  as 
color-octet contributions are suppressed by powers of the 
relative 
velocity 
$v$  
of the quark pair 
in the non-relativistic expansion. 
Near $z =1$, however, this power counting may be overcome 
due to the form of
 the short-distance 
coefficient associated with the color-octet component. 
This coefficient formally starts  
with a $\delta$-function distribution 
in lowest order. 
 The analysis of 
the hadronic smearing  in 
Ref.~[14]  
indicates that   
color-octet terms may become   
important  
within a range 
$\Delta z \sim v^2$ from the endpoint. 

 A crucial issue for the understanding of the endpoint spectrum 
 is the behavior of high-order terms in perturbation 
theory as $z \to 1$. It is known that potentially large 
contributions in $\ln (1 - z) $ 
may 
appear to all orders in $\alpha_s$, 
associated with the unbalance between real and 
virtual emission of soft gluons near the phase space boundary.  
   The problem then arises of 
summing these 
 logarithmically enhanced terms.  
In particular, 
it is important to see 
whether 
 this summation 
 gives rise to  
a Sudakov damping factor of the form  
$\sim \exp \left( - \alpha_s \ln^2 (1 - z) 
\right) $ in the 
decay width.  

This question is addressed in a perturbative analysis 
now in progress~\cite{chm}. This analysis shows that, in the 
color-singlet channel, 
the logarithms of $(1 - z)$ cancel  at each order in $\alpha_s$.  
 As a consequence,   
no Sudakov factor arises. In contrast, 
the color-octet channel does have a Sudakov  suppression. 

The reason 
for this behavior lies in the 
 properties of  
  coherence of color radiation.  
Qualitatively, 
the cancellation mechanism  
can be understood  from an illustration 
at one-loop level. Consider  
 the amplitude for the 
Born process in the singlet case, $\Upsilon (P) \to  
\gamma (k) + g (k_1) + g(k_2)$, and consider the emission of an 
additional soft gluon 
from this amplitude.  
In the 
leading infrared 
approximation, 
the decay width has the 
factorized 
structure 
\begin{equation}
\label{oneloop}
{{ d \Gamma} \over {  d z }} \sim 
{\bf J}^2 \, 
\left( {{ d \Gamma} \over {  d z }} \right)_{\rm {Born}}
\, d \Phi \hspace*{0.4 cm} , 
\end{equation}
where ${\bf J}$ is the eikonal current for soft gluon emission, and 
$d \Phi$ is the associated phase space.  
The standard power counting 
in terms of the soft gluon energy $\omega$ gives  
\begin{equation}
\label{phij}
d \Phi \sim \omega \, d \omega \, d \Omega \hspace*{0.2 cm} , 
\hspace*{0.4 cm} {\bf J}^2 \sim { 1 \over {\omega^2}} \, {{ k_1 \cdot k_2 }
\over { f (z ; {\mbox {angles}} )}} \hspace*{0.4 cm} ,  
\end{equation}
where $ d \Omega$ is the angular phase space, and $f$ is  a 
function of 
$z$ and 
the angles as $\omega \to 0$. 
Up to the first order in $\omega$ 
the gluon correlation is 
\begin{equation}
\label{k12}
 k_1 \cdot k_2 \sim M^2 \left[ (1 - z) + {\omega \over M} \, 
g (z ; {\mbox {angles}}) \right] \hspace*{0.4 cm} .
\end{equation}
That is, as $z \to 1$  
the photon recoils against two almost-collinear hard gluons. 
In this configuration, 
the logarithmic integration 
 $ d \omega \, / \, \omega $ in Eq.~(\ref{phij}) 
is canceled. One can show that 
this mechanism generalizes to all orders in $\alpha_s$. 

An interesting spin-off of this analysis is 
that 
the coherence scale 
for the QCD radiation 
associated with the 
decay 
is not 
the heavy quarkonium mass $M^2$ but rather the 
final-state 
invariant mass 
$(k_1+ k_2)^2$. 
Correspondingly, 
destructive interference occurs outside a cone  
with opening 
$\theta^2 \propto 1 - z$. 
  We note that, in contrast, the assumptions    
underlying the Monte Carlo model of Ref.~[8]  
amount to fixing the coherence scale to be $M^2$, 
and  $\theta^2 \sim 1 $.   
 This  model therefore does not take account of 
the gluon radiation in higher orders correctly. It would be 
of interest to see how  a 
  Monte Carlo model in which coherence  
 is correctly 
implemented compares with the  inclusive photon data.

\section*{Acknowledgments} 
 The  results  presented in this paper have been obtained in 
collaboration with  S. Catani and M. Mangano. 
This work is supported in part by the US Department of
Energy grant DE-FG03-96ER40969.

\end{document}